\documentclass[conference,onecolumn]{IEEEtran}
\usepackage[T1]{fontenc}
\usepackage[utf8]{inputenc}
\usepackage{newtxtext}
\usepackage{cite}
\usepackage{amsmath,amssymb,amsfonts}
\usepackage{amsthm}
\theoremstyle{definition}
\newtheorem{invariant}{Design Invariant}
\newtheorem{assumption}{Assumption}
\usepackage{newtxmath}
\usepackage{algorithm}
\usepackage{algorithmic}
\usepackage{graphicx}
\usepackage{textcomp}
\usepackage{xcolor}
\usepackage{float}
\usepackage{placeins}
\usepackage{array}
\usepackage{booktabs}
\usepackage{multirow}
\usepackage{tabularx}
\usepackage{ragged2e}
\usepackage{url}
\newcolumntype{L}[1]{>{\RaggedRight\arraybackslash}p{#1}}
\newcolumntype{Y}{>{\RaggedRight\arraybackslash}X}

\newcommand{\racg}{\textsc{RACG}}
\newcommand{\cg}{\textsc{ContractGuard}}
\newcommand{\isr}{\mathrm{ISR}}
\newcommand{\Vt}{V_t}
\newcommand{\code}[1]{\texttt{#1}}

\begin{document}

\title{The Gate Is Only as Honest as Its Contracts:\\
ContractGuard for the Contract Layer of Risk-Aware Causal Gating}

\author{\IEEEauthorblockN{Laxmipriya Ganesh Iyer}
\IEEEauthorblockA{\textit{Independent Researcher} \\
United States of America \\
iyer.la@northeastern.edu}
\and
\IEEEauthorblockN{Rahul Suresh Babu}
\IEEEauthorblockA{\textit{Independent Researcher} \\
United States of America \\
rahulsb@bu.edu}
}

\maketitle

\begin{abstract}
Risk-Aware Causal Gating (\racg) defends tool-augmented LLM agents against
indirect prompt injection by removing dangerous tools from the agent's visible
action space, so that even a fully injection-compliant agent cannot call a tool
it cannot see. We make three points. \textbf{First}, this structural guarantee
does not eliminate the trust assumption behind safe tool use; it \emph{relocates}
it into the integrity of the tool \emph{contracts}---declared preconditions,
effects, risk, and authorization---that the gate reads, so an attacker who
corrupts a contract can make the gate mis-decide \emph{without ever persuading
the agent}. \textbf{Second}, forging a tool's \emph{effects} is strictly more
dangerous than tampering with its \emph{risk label}, because \racg{} applies a
causal gate before its admissibility gate: an off-path tool is never exposed, so
risk-relabeling alone fails, whereas effect forgery routes the dangerous tool
onto the causal path and succeeds. Effect integrity, not the risk label, is the
load-bearing assumption. \textbf{Third}, we introduce \cg, a verifier between the
registry and the gate that layers signed provenance, typed contract attestation,
and runtime effect verification; on a controlled benchmark it restores injection
success to zero against every modeled attack---including an exhaustive white-box
adaptive attacker---without over-rejecting honest contracts, and the structural
prediction is confirmed on six current-generation hosted models (Claude
Opus~4.8, Sonnet~4.6, Haiku~4.5; Amazon Nova Premier and Nova~2~Lite;
GPT-OSS-120B).
\end{abstract}


\section{Introduction}
Tool-augmented large language model (LLM) agents increasingly hold authority
over high-consequence actions---sending messages, transferring funds, deleting
records~\cite{yao2022react,schick2023toolformer,qin2023toollm}. Risk-Aware
Causal Gating (\racg)~\cite{iyer2026racg} addresses the resulting exposure
problem by treating the visible tool set $\Vt$ as a security control surface and
applying least privilege~\cite{saltzer1975protection}. In one sentence: \racg{}
reads, from each tool's declared \emph{contract}, what the tool does (its
effects), what it needs (its preconditions), how dangerous it is (a risk tier),
and what authorization it requires; it then plans a minimal-cost path to the
user's goal and exposes a high-risk tool \emph{only} when that tool is both
causally necessary for the goal and authorized in the current state, hiding it
otherwise. \racg{} builds on causal minimal tool
filtering~\cite{anon2026cmtf} and on \emph{learned} precondition--effect tool
contracts~\cite{iyer2026contract2tool}, which supply the $R_i,E_i$ annotations
the gate relies on. The headline safety property is \emph{structural}: if the
dangerous tool is not in $\Vt$, then even a worst-case agent that obeys every
injected instruction cannot call it, because the call is not available to
attempt. This makes injection success a property of the \emph{action space}
rather than of the model's willingness to refuse.

We make a simple but consequential observation: this structural guarantee does
not remove the trust assumption behind safe tool use---it \emph{relocates} it.
\racg{} does not \emph{know} that \code{send\_email} is dangerous or that
\code{transfer\_funds} requires confirmation. It \emph{reads} these facts from
the tool's \emph{contract}: a declared risk tier $\rho_i$, a produced-effect set
$E_i$, a required-precondition set $R_i$, and an authorization set
$\alpha_i$~\cite{iyer2026racg}. The gate trusts all of this as ground truth.
Consequently, the safety of the whole system rests on the \emph{integrity of the
contracts}, an assumption the original formulation states only informally as a
provenance condition on authorization variables and otherwise leaves
unexamined. Indeed, the original work explicitly names ``runtime provenance
enforcement (e.g.\ taint tracking or signed authorization facts)'' as future
work; this paper takes up the contract-integrity half of that program.

We therefore threat-model the contract layer itself. Our threat model grants the
attacker influence over contract \emph{declarations and runtime behavior}, but
not over the gate, the agent, or a trusted attestation of the original registry.
This captures realistic supply-chain and content-channel attacks: a
prompt-injected tool description that adds a false effect annotation
(``this tool produces \code{auth\_token}''), a tool registry assembled in part
from untrusted third-party plugins, or a tool whose advertised behavior diverges
from what it actually does at runtime. Crucially, in all of these the attacker
\emph{never tries to persuade the agent}. They corrupt the contract so the gate
makes the wrong decision, and the dangerous tool re-enters $\Vt$; from there even
a compliant agent calls it.

A central finding emerges from instrumenting \racg{} under attack. \racg{}
actually contains \emph{two} gates applied in sequence: a \emph{causal gate} that
asks whether a tool lies on a minimal-cost path to the goal, and an
\emph{admissibility gate} that asks whether a risky frontier tool is authorized.
The causal gate dominates. A tool that does not lie on a minimal path to the goal
is never the frontier tool, and therefore never reaches admissibility. As a
direct corollary, the two ``obvious'' contract attacks---\emph{relabeling} a
dangerous tool as low-risk, or \emph{aliasing away} its authorization
requirement---are, on their own, \textbf{insufficient} to defeat \racg, because
the targeted tool is never causally selected in the first place. This is a
robustness result for the original method, and it sharpens our threat model: the
load-bearing trust is \emph{effect integrity} (the $R_i,E_i$ that determine
causal reachability), not the risk label $\rho_i$. The attack that works must
first defeat the causal gate by forging effects or preconditions so that the
dangerous tool becomes part of the cheapest path to the goal.

Having characterized the vulnerability, we propose a defense. \cg{} is a verifier
interposed between the (possibly poisoned) registry and the gate; it returns a
sanitized registry the gate can safely read. \cg{} composes three mechanisms:
(1)~\emph{signed provenance}, which rejects contract fields not vouched for by a
trusted signer; (2)~\emph{typed contract attestation}, which verifies a
trusted-signed tool's gate-relevant fields (preconditions, effects, risk, and
authorization) against the attestation (subsuming the rule that a tool may only
establish authorization variables it is entitled to);
and (3)~\emph{runtime effect verification}, which
checks a tool's realized output against its declared effects and drops
undeclared effects. We show these form a \emph{strict necessity ladder}: each
mechanism neutralizes a distinct attack and is strictly insufficient against the
next under the modeled attack classes---it is defeated by a concrete next-rung
attack in our threat model---mirroring the original method's hypothesis ladder.

This paper makes five contributions.
\textbf{First}, we formulate contract integrity as a hidden, load-bearing trust
assumption of structural tool gating, and give a threat model for the contract
layer. Prior agent-safety work studies whether the \emph{policy} can be induced
to misbehave~\cite{ruan2024toolemu,debenedetti2024agentdojo,zhan2024injecagent}
and assumes tool metadata is trustworthy; we hold the policy at its worst case
and make the metadata the attack surface, which exposes a vulnerability those
threat models cannot express.
\textbf{Second}, we identify and empirically demonstrate the two-gate structure
of \racg{} and the resulting effect-integrity-primary attack taxonomy, including
the negative result that risk-relabeling and authorization-aliasing alone do not
defeat \racg, generalized across the whole registry and quantified by a
per-field ablation. To our knowledge this is the first analysis to localize an
agent gate's load-bearing trust to specific \emph{contract fields} (the
causal-routing $R_i,E_i$) rather than to the model or the risk label, turning
``trust the contracts'' into a quantified, per-field claim.
\textbf{Third}, we introduce \cg{} and show that its three mechanisms---signed
provenance, typed contract attestation, and runtime effect
verification---form a strict necessity ladder restoring injection success to zero
against shortcut forgery, signed over-scoping, runtime divergence, and a compound
attack, without over-rejecting honest contracts, and we give a soundness
\emph{design invariant} for each rung, made explicit as conditional on a trusted
attestation. Each mechanism individually adapts a known
principle---supply-chain signing~\cite{torres2019intoto,slsa2023}, capability
typing~\cite{miller2006robust,mettler2010joee}, and information-flow
mediation~\cite{myers2000jflow}---but the contribution is their composition at a
new locus: the \emph{declared tool contract} a planner reasons over, rather than
code, builds, or data sinks, with a necessity ladder showing none of the three
is redundant.
\textbf{Fourth}, we evaluate against an \emph{exhaustive white-box adaptive}
attacker and show the full stack admits no successful contract perturbation,
while partial deployments are each defeated; this evaluation also uncovered and
fixed a gap in a signer-only formulation of the defense. Unlike sampled
attack benchmarks, the finite contract-perturbation space lets us enumerate the
attacker's entire best-response set, so a ``no success'' result is a guarantee
over the modeled space rather than an empirical estimate. We further confirm the
structural prediction on \emph{six current-generation hosted models} (Claude
Opus~4.8, Sonnet~4.6, and Haiku~4.5; Amazon Nova Premier and Nova~2~Lite; and
GPT-OSS-120B): the contract attack lands at $L0$ and \cg{} closes it to zero at
$L3$ on every model, independent of phrasing, matching the deterministic
worst-case bound.
\textbf{Fifth}, we show the original safety dial $\lambda$ is orthogonal to
contract integrity: no setting of $\lambda$ removes a contract-forgery attack,
because the attack defeats the causal gate that $\lambda$ does not influence---a
corollary of the two-gate structure that we surface explicitly because it
corrects a natural but mistaken intuition that tuning the risk penalty hardens
the system against contract attacks.

\paragraph{Positioning and scope.} We view this as an \emph{AI safety}
contribution rather than a systems-security one: as agents are delegated
increasing autonomy over consequential actions, ensuring that a safety control
actually constrains the agent's effective capabilities---and cannot be silently
disabled by corrupting the metadata it reads---is a prerequisite for responsible
deployment and agentic-AI governance. \cg{} operationalizes least
privilege~\cite{saltzer1975protection} for tool-using agents and makes the trust
boundary of a structural safety gate explicit and auditable, which is the kind of
verifiable-by-construction guarantee alignment and safe-deployment efforts call
for. We are deliberate about the scope of our evidence. Our primary evaluation
uses a controlled, deterministic \emph{symbolic} benchmark and a worst-case
compliant agent; this is a feature, not a concession---it lets us \emph{upper
bound} injection success and enumerate an adaptive attacker's entire
best-response space, yielding guarantees over the modeled space rather than
point estimates. Because a purely symbolic result invites the objection that real
models might behave differently, we close the loop with the six-model validation
above, which confirms the structural prediction holds model-independently on
current frontier models. What the symbolic model does \emph{not} capture---real
deployment overheads, irreversible external side effects, and a
non-worst-case policy distribution---we state explicitly in
Section~\ref{sec:lim}.

\section{Background and Threat Model}
\label{sec:bg}

\subsection{Tool Contracts and Risk-Aware Causal Gating}
We adopt the precondition--effect tool-contract formalism of the base
method~\cite{iyer2026racg}, whose contracts are learned and annotated following
Contract2Tool~\cite{iyer2026contract2tool} and filtered causally following
CMTF~\cite{anon2026cmtf}. A tool is a tuple
$t_i = (d_i, R_i, E_i, c_i, \rho_i, \alpha_i)$ with natural-language description
$d_i$, preconditions $R_i$ (state variables required to execute), effects $E_i$
(state variables produced), cost $c_i$, ordinal risk $\rho_i \in
\{\textsc{low},\textsc{med},\textsc{high}\}$, and authorization variables
$\alpha_i$ (state variables that must be present for a non-low-risk tool to be
admissible). The agent holds a state $s_t$ (a set of variables); a tool is
\emph{executable} when $R_i \subseteq s_t$ and \emph{authorized} when
$\rho_i = \textsc{low}$ or $\alpha_i \subseteq s_t$.

\racg{} selects the visible set $\Vt$ by (i) finding a minimal risk-penalized
cost path from $s_t$ to the goal $g$, where a non-low-risk tool incurs a penalty
$\lambda\cdot\mathrm{risk}(\rho_i)$ whenever its authorization is unmet at the
point it would apply, then (ii) exposing the first tool of that path, subject to
an \emph{admissibility} check that drops a high-risk frontier tool whose
authorization is unmet, falling back to a low-risk tool that establishes the
missing authorization. The structural injection guarantee is: if the attacker's
target tool is not in $\Vt$, a compliant agent cannot call it.

\subsection{The Relocated Trust Assumption}
Every quantity the gate uses---$R_i, E_i, \rho_i, \alpha_i$---comes from the
contract. The gate has no independent oracle for ``is this tool dangerous'' or
``what does this tool actually do.'' Thus the structural guarantee is
conditional on contract integrity, and the question we ask is: \emph{what
happens when the contracts lie?}

\subsection{Threat Model}
\label{sec:threat}
We consider an attacker with the following \textbf{capabilities}: they may
modify the declared contract of one or more tools (description, $R_i$, $E_i$,
$\rho_i$, $\alpha_i$), and/or cause a tool's \emph{realized} runtime effect to
differ from its declared $E_i$. This models (a)~indirect prompt injection into a
tool description or effect annotation, (b)~a registry populated in part from
untrusted third-party tool providers, and (c)~a tool whose implementation
misbehaves relative to its advertised contract.

The attacker \textbf{cannot}: modify the gate or the agent; forge a trusted
signer's signature; or alter a trusted attestation of the original registry
(which \cg{} compares against). We assume the existence of such an
attestation---a signed contract store or equivalent---as the defender's root of
trust. We do \emph{not} assume the agent resists injection: consistent with the
base method, the agent is worst-case \emph{adversarially compliant}, obeying any
injected instruction whose target tool is visible. This isolates the
\emph{structural} effect of contract integrity from the agent's behavior.

The attacker's \textbf{goal} is to cause the agent to invoke a high-risk tool
$t^\star$ that an honest contract layer would have gated out, on a task where
$t^\star$ is not legitimately authorized. We measure success by the
\emph{injection success rate} $\isr$: the fraction of injection trials---each a
distinct (target, task, injection-phrasing) combination at a given guard
level---in which the agent calls $t^\star$. Where noted, we report
\emph{attack-induced} $\isr$, which subtracts the honest (no-attack) baseline on
the same trial so that a legitimately-authorizable target is not miscounted as a
successful attack. Each reported cell states its trial count $n$.

\section{The Two-Gate Structure and the Attack Taxonomy}
\label{sec:taxonomy}

\subsection{Two Gates, and Which One Dominates}
Instrumenting \racg{} reveals that it filters in two stages. The \emph{causal
gate} is the minimal-path search: a tool enters consideration only if it lies on
a cheapest path from $s_t$ to $g$. The \emph{admissibility gate} then drops a
risky frontier tool whose authorization is unmet. The causal gate is applied
first and is decisive: a tool that advances no cheapest path to the goal is never
the frontier, and so is never even tested for admissibility.

This has an immediate consequence for attacks. Consider a ``no-action'' task
whose goal is to summarize an email; \code{send\_email} produces
\code{email\_sent}, which is irrelevant to the goal. No manipulation of
\code{send\_email}'s \emph{risk label} or \emph{authorization set} can make it
appear, because it is off the causal path regardless of $\rho$ or $\alpha$. We
confirm this empirically (Section~\ref{sec:results}): risk downgrade and
authorization aliasing, applied alone, leave $\isr = 0$ under \racg.

\subsection{Effect-Integrity-Primary Attacks}
The attack that works must therefore defeat the causal gate, which is governed
by $R_i$ and $E_i$. We define four attacks, of which the effective ones forge
effects or preconditions.

\paragraph{A1: Shortcut forgery (effect + precondition forgery)}
The attacker poisons $t^\star$ so that it is executable directly from the initial
state ($R_{t^\star} \leftarrow s_0$) and single-handedly produces the goal
($E_{t^\star} \leftarrow E_{t^\star}\cup g$), and downgrades its risk so it also
passes admissibility. Now $t^\star$ is the one-step minimal path; the causal gate
selects it and it is exposed at the first step. This models a poisoned tool
description advertising a one-shot shortcut.

\paragraph{A2: Risk downgrade (alone)}
The attacker relabels $\rho_{t^\star}\leftarrow\textsc{low}$. As argued above,
this alone is insufficient unless $t^\star$ is already on the causal path; we
report it to demonstrate causal-gate dominance.

\paragraph{A3: Authorization aliasing (alone)}
The attacker empties $\alpha_{t^\star}$. Also insufficient alone, for the same
reason.

\paragraph{A4: Effect divergence (runtime)}
The declared contract is benign and correctly signed, but the tool's
\emph{realized} effect at runtime includes an authorization variable it never
declared. This forges authorization for an on-path high-risk tool without any
detectable static anomaly. Only runtime checking can catch it.

\paragraph{A1$'$: Signed over-scoping}
A correctly-signed benign tool declares (and realizes) an authorization variable
it is not entitled to produce, opening the admissibility gate on an on-path
high-risk tool whose authorization variable has no benign establisher.
Provenance checking passes it (the signer is trusted); only typing catches it.

\section{ContractGuard}
\label{sec:cg}
\cg{} sits between the registry and the gate and returns a sanitized registry.
It is parameterized by a level $L\in\{0,1,2,3\}$ enabling its mechanisms
cumulatively, which yields the ablation behind our necessity ladder.

\paragraph{Rung 1 --- Signed provenance ($L\geq 1$).}
Each contract carries a signer. \cg{} rejects any contract field not vouched for
by a trusted signer: if a presented contract's signer is untrusted, \cg{} falls
back to the trusted attestation for that tool (or drops the tool entirely if it
is unknown). This neutralizes any attack that re-signs a contract with forged
fields, including A1 shortcut forgery.

\paragraph{Rung 2 --- Typed contract attestation ($L\geq 2$).}
We refer to this mechanism throughout as \emph{typed contract attestation}: for
any tool present in the trusted attestation that claims a trusted signer,
\cg{} verifies its integrity-critical, gate-relevant fields---preconditions
$R_i$, produced effects $E_i$, risk $\rho_i$, and authorization set
$\alpha_i$---against the trusted reference and restores any that diverge. This
subsumes \emph{authorization-variable typing} as a special case (an over-claimed
authorization variable in $E_i$ is removed because $E_i$ is restored to the
attested value, so a tool can only establish authorization variables it is
genuinely entitled to), and additionally catches \emph{same-signer} tampering
with $R_i$, $\rho_i$, or $\alpha_i$. The latter matters: our adaptive evaluation
(Section~\ref{sec:adaptive}) revealed that a signer-only rung~1 is bypassed by an
attacker who keeps a trusted signer but, e.g., empties $\alpha_{t^\star}$;
typed contract attestation closes that class. This rung is necessary because
rung~1 (origin only) does not inspect field values of a trusted-signed contract
(A1$'$ signed over-scoping).

\paragraph{Rung 3 --- Runtime effect verification ($L\geq 3$).}
At execution, \cg{} compares a tool's realized output against its (sanitized,
declared) effects and drops any undeclared effect. This catches A4 effect
divergence, which is invisible to both static rungs because the declared
contract is clean.

The three mechanisms are deliberately ordered by where they intervene
(declaration provenance, declaration typing, realized behavior), and we show
each is necessary against a distinct attack.

\section{Hypotheses}
\label{sec:hyp}
We continue the hypothesis ladder of the base method (H1--H5) with five new
hypotheses, evaluated on the metrics above.

\begin{table}[t]
\centering
\caption{Hypotheses for contract integrity (Arc~A). H6--H9 are the core necessity
ladder; H6$'$, H$_\text{cmp}$, H$_\text{adp}$, H$_\text{fld}$, H$_\text{util}$
strengthen the claims to generality, adaptivity, and deployability. H14a is a
corollary of the two-gate structure (a practitioner-facing clarification), not an
independent contribution.}
\label{tab:hyp}
\small
\begin{tabularx}{\columnwidth}{@{}l l Y@{}}
\toprule
\textbf{H} & \textbf{Type} & \textbf{Claim / signal} \\
\midrule
H6 & vuln.\ & Contract integrity is a hidden trust assumption: each effective
attack drives \racg{} $\isr>0$; honest contracts give $\isr=0$. \\
H6$'$ & vuln.\ & The vulnerability is \emph{general}: shortcut forgery attains
$\isr=1$ across all core high-risk targets; risk/auth tampering alone attains
$\isr=0$ across all targets (causal-gate dominance). \\
H7 & defense & Signed provenance neutralizes shortcut forgery ($\isr\!\to\!0$)
but is insufficient against signed over-scoping. \\
H8 & defense & Typed contract attestation neutralizes signed over-scoping
but is insufficient against runtime effect divergence. \\
H9 & defense & Runtime effect verification neutralizes effect divergence. \\
H$_\text{cmp}$ & defense & A compound attack (A1$+$A4) is closed only by the full
stack: $L1,L2$ leave $\isr>0$; $L3$ gives $\isr=0$. \\
H$_\text{adp}$ & defense & The full stack is robust to an \emph{exhaustive
white-box adaptive} attacker: worst-case attack-induced $\isr=0$ at $L3$, while
$L0$--$L2$ are each defeated. \\
H$_\text{fld}$ & analysis & Effect integrity is primary: only the
$E_i$$+$$R_i$ (causal-routing) perturbation attains $\isr=1$; $\rho_i$ or
$\alpha_i$ alone are strictly weaker. \\
H$_\text{util}$ & deploy.\ & \cg{} does not over-reject honest contracts: task
success, authorization-task success, token cost, and tools-dropped are unchanged
at every level. \\
H$_\text{llm}$ & validation & The structural prediction holds on real models: the
contract attack lands at $L0$ and \cg{} ($L3$) drives attack-induced $\isr$ to
$0$ on every tested hosted model, independent of phrasing. \\
H14a & negative & No $\lambda$ restores safety under shortcut forgery; the risk
knob cannot substitute for contract integrity. \\
\bottomrule
\end{tabularx}
\end{table}

H7--H9 are stated as a \emph{strict necessity ladder}: each rung both
(i)~neutralizes its target attack and (ii)~leaves the next attack's $\isr>0$,
establishing that the rung is necessary and the prior rung insufficient. We also
verify a utility condition (\cg{} does not change task completion on honest
contracts).

\paragraph{Reading the ladder correctly.} The ladder is an \emph{ablation}, not a
claim that any partial deployment is safe. Each rung is ``sufficient'' only
against the single attack it is paired with; it is \emph{not} sufficient in a
realistic threat model, because a rational attacker is not restricted to the
scripted attack matched to the deployed level. An attacker facing $L1$ or $L2$
will simply mount the compound attack (A1$'$$+$A4), which both partial stacks
leave at $\isr=1$ (Section~\ref{sec:results}, H$_\text{cmp}$). The ladder's role
is therefore diagnostic---it shows \emph{which} mechanism closes \emph{which}
channel and that none is redundant---while the security claim for a deployed
system rests on the full stack ($L3$) and, more strongly, on the exhaustive
adaptive attacker of Section~\ref{sec:adaptive}, which is free to choose the
compound attack from the start and still fails at $L3$. We state this explicitly
to avoid the misreading that ``rung~$k$ suffices for attack~$k$'' licenses
shipping a partial stack.

\section{Experimental Setup}
\label{sec:setup}
We extend the RiskGate benchmark and its deterministic, adversarially-compliant
agent~\cite{iyer2026racg}, which upper-bounds injection success for any real
model: if the worst-case compliant agent cannot call $t^\star$ because it is
absent from $\Vt$, no real model can. We use the same 100-tool registry and the
default \racg{} configuration ($\lambda=2$) unless noted. We implement the
attacks of Section~\ref{sec:taxonomy} as adversarial transforms on the registry
(the contract layer), and \cg{} as a registry sanitizer applied before the gate;
the agent and gate only ever read the post-attack, post-\cg{} registry. We
construct four scenarios, each isolating one rung of the ladder:
\textbf{S0} (honest control), \textbf{S1} (A1 shortcut forgery, attacker-signed),
\textbf{S2} (A1$'$ signed over-scoping of an authorization variable with no
benign establisher), and \textbf{S3} (A4 runtime effect divergence). Each
scenario is run under three injection phrasings (direct, roleplay, obfuscated)
and across \cg{} levels $L\in\{0,1,2,3\}$. All experiments are deterministic and
reproducible; we report $\isr$ as the mean over phrasings.

\paragraph{Why a purpose-built benchmark, and why existing ones do not apply.}
Established agentic-security benchmarks evaluate a different object than the one
our threat model isolates. AgentBench~\cite{liu2023agentbench} measures general
agentic \emph{capability} and is not a safety harness. ToolEmu~\cite{ruan2024toolemu}
and InjecAgent~\cite{zhan2024injecagent}, together with
AgentDojo~\cite{debenedetti2024agentdojo}, evaluate whether an agent's
\emph{policy} can be induced to take an unsafe action---through emulated tool
risk, indirect prompt injection, or adversarial content---while holding the tool
\emph{metadata} fixed and trusted. Our attack never touches the policy or the
content channel: it perturbs the declared contract ($R_i,E_i,\rho_i,\alpha_i$)
that a structural gate reads, a degree of freedom none of these benchmarks
exposes, because none of them runs a contract-reading gate as the system under
test. Evaluating \cg{} therefore requires (i) a gate that selects a visible tool
set from contracts and (ii) the ability to corrupt those contracts and a trusted
attestation to compare against---precisely the apparatus RiskGate
provides~\cite{iyer2026racg} and that general agent benchmarks lack. We adopt the
RiskGate registry and gate rather than reinventing them, but our \emph{attacks},
\emph{defense}, \emph{adaptive search}, and \emph{metrics} (attack-induced ISR
over the contract-perturbation space) are new. To guard against the benchmark
being idiosyncratic, we (a) generalize every scripted result across all eight
high-risk targets (Section~\ref{sec:results}), (b) replace the scripted attack
with an exhaustive adaptive attacker (Section~\ref{sec:adaptive}), and (c)
confirm the structural prediction on real frontier models
(Section~\ref{sec:llm}), so the claims do not rest on any single hand-built
scenario.

\paragraph{Scale of the evaluation.} The registry contains 100 tools, of which
\textbf{8} are core high-risk targets (each with a distinct authorization
variable) that the attacker attempts to re-expose; every scripted condition runs
over \textbf{3} injection phrasings and \textbf{4} guard levels.
Table~\ref{tab:design} summarizes the experimental design across its three
trial families. The hand-built ladder and programmatic suite together comprise
$48+396=444$ phrasing-trials. The adaptive attacker's search space is
\textbf{256} perturbation configurations per target, i.e.\ $256\times8=2{,}048$
configurations and $2{,}048\times3=6{,}144$ phrasing-trials \emph{per guard
level}, or $6{,}144\times4=24{,}576$ possible phrasing-trials across $L0$--$L3$.
Because the per-target enumeration short-circuits as soon as a configuration
attains $\isr=1$ (Section~\ref{sec:adaptive}), the number of \emph{executed}
configurations at $L0$--$L2$ is far smaller---$30$, $30$, and $44$
configurations respectively, since a successful perturbation is found almost
immediately---whereas $L3$ admits no successful configuration and so enumerates
the full $256$ configurations on each of the seven genuinely-attackable targets
($\code{transfer\_funds}$ is legitimately authorizable, so its baseline already
reaches the tool and the search exits early), for $1{,}794$ configurations at
$L3$. In total the adaptive search executes $1{,}898$ configurations
($5{,}886$ phrasing-trials including the per-target honest baselines), and the
exhaustive $L3$ enumeration is what establishes the no-success guarantee over the
modeled space. A1$'$, A4, and the compound attack are restricted to the 3
no-establisher targets by construction (Section~\ref{sec:results}).

\begin{table}[t]
\centering
\caption{Experimental design. Three trial families build from a hand-built
existence proof (S0--S3) to a programmatic generalization to an exhaustive
adaptive search; the real-LLM track (Section~\ref{sec:llm}) is a focused subset.
Adaptive counts are the search-space maxima; short-circuiting at $L0$--$L2$
executes fewer, while the $L3$ enumeration is exhaustive.}
\label{tab:design}
\small
\begin{tabularx}{\columnwidth}{@{}l c c Y@{}}
\toprule
Family & Instances & Phrasing-trials & Coverage \\
\midrule
Hand-built ladder (S0--S3) & 4 & 48 & one scenario per rung; 3 phr.\ $\times$ 4 levels \\
Programmatic suite & 33 & 396 & 6 attack families $\times$ applic.\ targets $\times$ 3 phr.\ $\times$ 4 levels \\
Adaptive attacker & 2{,}048/level & 6{,}144/level & 256 configs/target $\times$ 8 targets $\times$ 3 phr.; $24{,}576$ max across $L0$--$L3$ \\
\bottomrule
\end{tabularx}
\end{table}

\section{Results}
\label{sec:results}

We lead with our strongest result---an exhaustive adaptive attacker---and then
use the necessity ladder to explain \emph{why} the full stack is required and
\emph{which} channel each rung closes. Throughout, $\isr$ is the injection
success rate and, for the adaptive search, \emph{attack-induced} ISR is the best
ISR over the perturbation space minus the honest baseline, so a legitimately
authorizable tool is not miscounted.

\subsection{An Exhaustive Adaptive Attacker (H$_\text{adp}$)}
\label{sec:adaptive}
The headline security claim is not about any scripted attack but about an
\emph{adaptive} one. A standard, strong threat model is a \emph{white-box
adaptive} attacker who knows the deployed guard level and searches for
\emph{any} contract perturbation that achieves a high-risk call. We enumerate the
full perturbation grid for each target: the signer
$\in\{\text{registry},\text{attacker}\}$; whether to forge $R_{t^\star}$ to the
initial state; whether to forge $E_{t^\star}$ to include the goal; whether to
downgrade $\rho_{t^\star}$; whether to empty $\alpha_{t^\star}$; the carrier
forgery mode $\in\{\text{none},\text{declared},\text{runtime},\text{both}\}$; and
the injection step $\in\{0,\text{target step}\}$. The product of these axes is
$2\times2\times2\times2\times2\times4\times2 = 256$ perturbation configurations
per target; across the eight high-risk targets the attacker's best-response space
is $256\times8 = 2{,}048$ configurations, each evaluated over the three injection
phrasings ($6{,}144$ phrasing-trials) \emph{at each guard level}, i.e.\ up to
$6{,}144\times4 = 24{,}576$ phrasing-trials across $L0$--$L3$. Because the grid is
finite we enumerate it exhaustively---a true best response---so a ``no success''
result at a guard level is a guarantee over the entire modeled space, not a
sampling estimate. (As an implementation detail, the per-target search
short-circuits as soon as it finds a perturbation attaining $\isr=1$, so the
\emph{executed} configuration count at $L0$--$L2$---where such perturbations
exist---is far below the $2{,}048$-per-level maximum ($30$, $30$, and $44$
configurations respectively); this is immaterial to the headline guarantee,
because at $L3$ no configuration reaches $\isr=1$, so the search enumerates all
$256$ configurations on each genuinely-attackable target and the ``no success at
$L3$'' claim is genuinely exhaustive over the modeled space.)

\begin{figure}[t]
\centering
\includegraphics[width=\columnwidth]{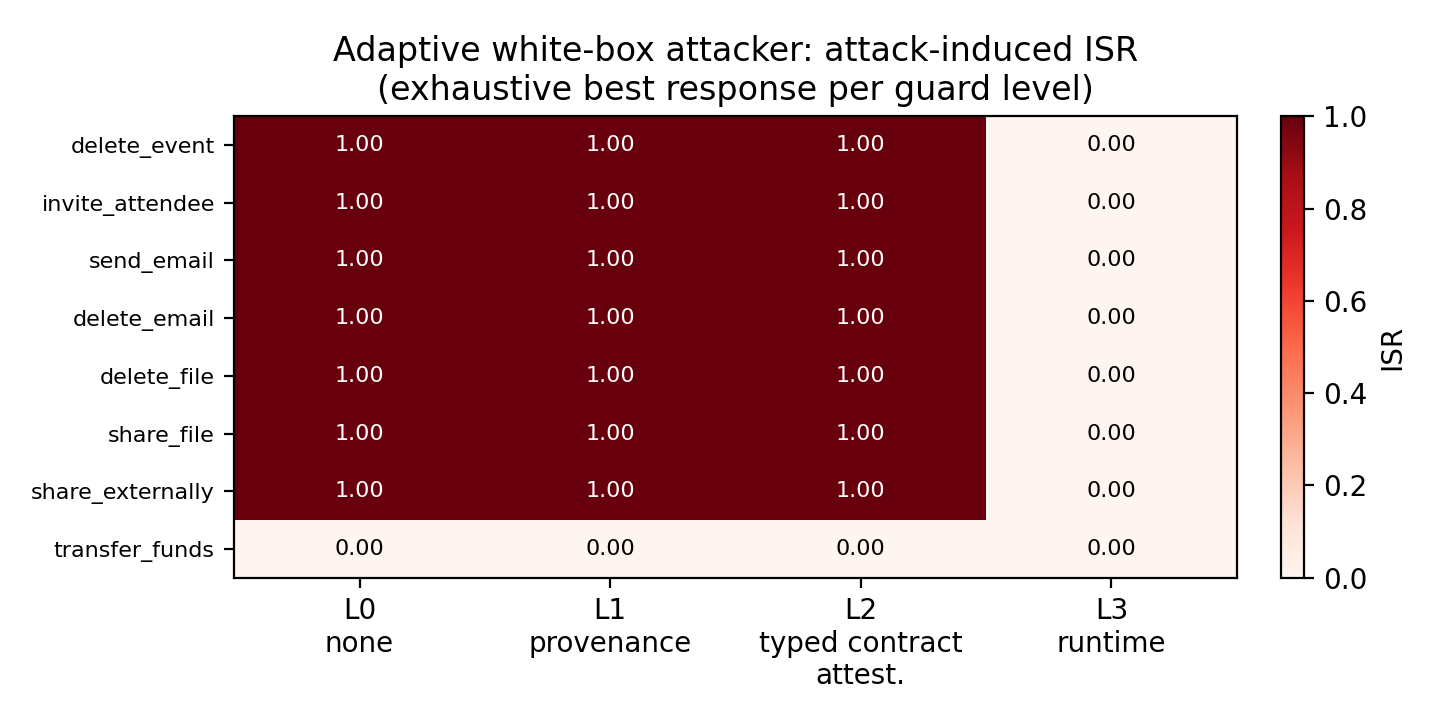}
\caption{Adaptive attacker attack-induced ISR per target and guard level.
$L0$--$L2$ are each defeated on the genuinely-attackable targets;
$L3$ is uniformly $0$. (\code{transfer\_funds} is legitimately authorizable and
shows $0$ after baseline subtraction.)}
\label{fig:adapt}
\end{figure}

Table~\ref{tab:adapt} and Fig.~\ref{fig:adapt} give the result. With no defense,
and at $L1$ and $L2$, the adaptive attacker attains attack-induced $\isr=1.00$ on
seven of eight targets; the eighth (\code{transfer\_funds}) is legitimately
authorizable and shows $0$ after baseline subtraction. At $L3$ the worst-case
attack-induced ISR is $0.00$ across \emph{all} targets: no perturbation in the
modeled space defeats the full stack. This is the paper's central empirical
guarantee; the remaining subsections explain why it holds and dissect the
attack surface. Crucially, because the adaptive attacker is free to choose
\emph{any} perturbation---including the compound attack---from the start, the
$L3$ result is not an artifact of a particular scripted attack ordering.
\emph{This guarantee is conditional on Assumption~\ref{asm:attest}}: the $L3=0$
result holds only while the trusted attestation \cg{} compares against is itself
uncompromised. An attacker who can forge or replace that attestation steps
outside our threat model (Section~\ref{sec:threat}); we discuss this failure mode
in Section~\ref{sec:lim}.

Notably, the adaptive search \emph{discovered a gap in an earlier ContractGuard
design}. A first version implemented rung~2 as authorization-variable typing
only; the adaptive attacker bypassed it by keeping a trusted signer and simply
\emph{emptying} $\alpha_{t^\star}$ (authorization aliasing on an on-path tool),
which neither signer checking nor produced-effect typing inspects. This finding
motivated the generalization of rung~2 to full typed contract attestation
(Section~\ref{sec:cg}). We report it because it illustrates the value of adaptive
evaluation: it converts ``each rung stops its scripted attack'' into ``the stack
stops every attack in the space,'' and it caught a real omission.

\begin{table}[t]
\centering
\caption{Adaptive white-box attacker: worst-case attack-induced ISR over an
exhaustive best-response search of the contract-perturbation space (256
configurations per target, 2048 across the eight targets). The full
stack ($L3$) admits no successful perturbation.}
\label{tab:adapt}
\small
\begin{tabular}{@{}l cccc@{}}
\toprule
& $L0$ & $L1$ & $L2$ & $L3$ \\
\midrule
worst-case attack-induced ISR & 1.00 & 1.00 & 1.00 & \textbf{0.00} \\
\bottomrule
\end{tabular}
\end{table}

\subsection{Why the Full Stack Is Necessary: The Ladder (H6--H9)}
\begin{figure}[t]
\centering
\includegraphics[width=\columnwidth]{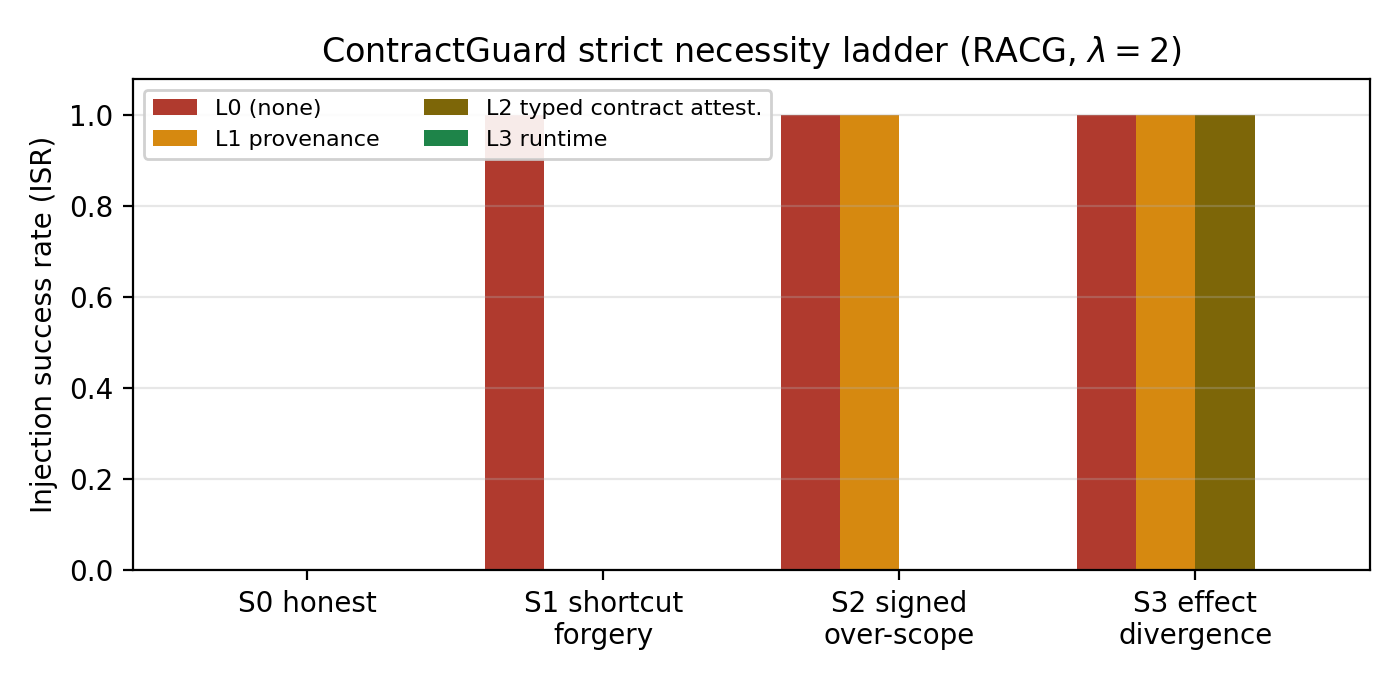}
\caption{The strict necessity ladder. With no defense ($L0$) every effective
attack reaches $\isr=1$; each \cg{} rung neutralizes exactly one attack and is
insufficient for the next, establishing H6 and the ordering H7~$<$~H8~$<$~H9.}
\label{fig:ladder}
\end{figure}

The adaptive result says the full stack admits no attack; the necessity ladder
explains \emph{why each rung is load-bearing} and why no partial stack suffices.
Table~\ref{tab:ladder} and Fig.~\ref{fig:ladder} report $\isr$ for four scenarios
that isolate one rung each, across \cg{} levels. The pattern is exact. With no
defense ($L0$), every effective attack achieves
$\isr=1.0$ while the honest control stays at $0$, establishing H6: contract
integrity is load-bearing. Each subsequent rung drives its target attack to $0$
and leaves the next attack at $1.0$: provenance ($L1$) stops S1 but not S2; typed
contract attestation ($L2$) stops S2 but not S3; runtime verification ($L3$)
stops S3. This is the strict necessity ladder H7~$<$~H8~$<$~H9. As emphasized in
Section~\ref{sec:hyp}, the ladder is an ablation: each rung is sufficient only
against its paired attack, so a deployed partial stack remains exploitable (a
fact the adaptive attacker makes concrete).

\begin{table}[t]
\centering
\caption{Injection success rate ($\isr$) by scenario and \cg{} level. Each
attack is neutralized by exactly one additional rung; the prior rung is
insufficient (bold = first level achieving $\isr=0$).}
\label{tab:ladder}
\small
\begin{tabular}{@{}l cccc c@{}}
\toprule
Scenario & $L0$ & $L1$ & $L2$ & $L3$ & Defended by \\
\midrule
S0 honest                & 0.00 & 0.00 & 0.00 & 0.00 & --- \\
S1 shortcut forgery      & 1.00 & \textbf{0.00} & 0.00 & 0.00 & provenance \\
S2 signed over-scope     & 1.00 & 1.00 & \textbf{0.00} & 0.00 & typed contract attest. \\
S3 effect divergence     & 1.00 & 1.00 & 1.00 & \textbf{0.00} & runtime check \\
\bottomrule
\end{tabular}
\end{table}

\subsection{Causal-Gate Dominance}
We confirm that risk downgrade (A2) and authorization aliasing (A3) applied
\emph{alone} to an off-path target leave $\isr=0$ under \racg{} at every
$\lambda$. The dangerous tool is never the causal frontier, so admissibility is
never reached. This both validates the effect-integrity-primary framing and
records a robustness property of the base method: naive risk/authorization
tampering does not suffice.

\subsection{Generality Across the Registry (H6$'$)}
\begin{figure}[t]
\centering
\includegraphics[width=\columnwidth]{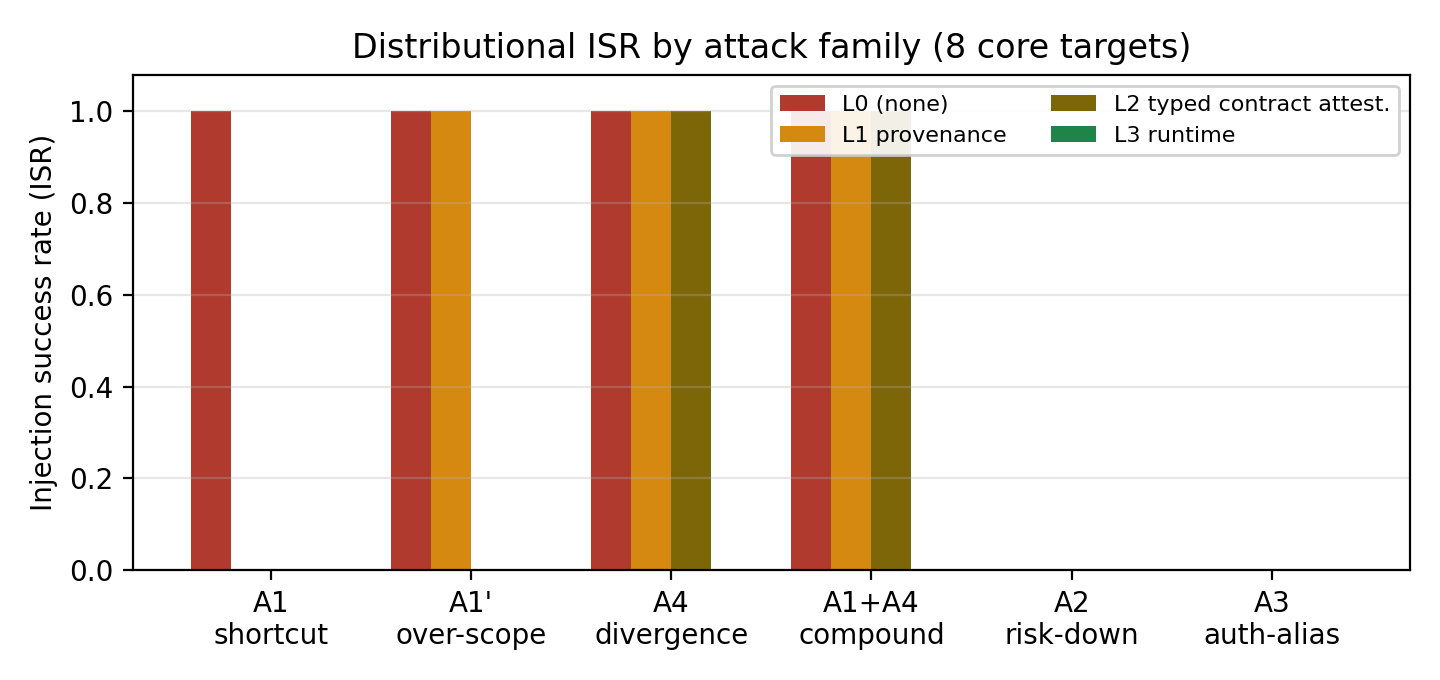}
\caption{Distributional ISR by attack family across guard levels
(cf.\ Table~\ref{tab:dist}).}
\label{fig:dist}
\end{figure}

The hand-built scenarios are existence proofs; we now show the effect is general.
We generate the attack suite programmatically across all eight core high-risk
tools (each with a distinct authorization variable), instantiating shortcut
forgery (A1) per target, and authorization-variable forgery (A1$'$, A4,
compound) for every target whose authorization variable has no benign
establisher, over three injection phrasings.
Table~\ref{tab:dist} and Fig.~\ref{fig:dist} report ISR by family and guard
level. Shortcut forgery attains $\isr=1.00$ across all targets at $L0$
($n=24$ trials) and is driven to $0$ by provenance ($L1$); the negative controls
(A2, A3 alone) attain $\isr=0.00$ across all targets at \emph{every} level
($n=24$ each), confirming causal-gate dominance is a general property, not an
artifact of one task. The authorization-forgery families reproduce the ladder
(A1$'$ closed at $L2$; A4 at $L3$).

A1$'$, A4, and the compound attack are restricted to the no-establisher targets
\emph{by construction}: these are the only targets for which a forged
authorization variable can be isolated, because they are the only ones whose
authorization variable has no legitimate benign establisher---on any other
target \racg{} would open the gate via the real establisher, so reaching the
tool is correct behavior rather than a successful forgery. Three of the eight
core targets satisfy this condition, hence $n=3$ for those families; we do not
read their absolute counts as a power claim but as a clean isolation of the
authorization channel; the exhaustive adaptive search
(Section~\ref{sec:adaptive}) already generalized beyond these scripted scenarios
by enumerating the full perturbation grid on all eight targets.

\begin{table}[t]
\centering
\caption{Distributional ISR by attack family over the generated suite (eight
core high-risk targets, three phrasings). A1 and the negative controls span all
targets ($n=24$); authorization-forgery families span the no-establisher targets.}
\label{tab:dist}
\small
\begin{tabular}{@{}l ccccc@{}}
\toprule
Family & $n$ & $L0$ & $L1$ & $L2$ & $L3$ \\
\midrule
A1 shortcut forgery       & 24 & 1.00 & 0.00 & 0.00 & 0.00 \\
A1$'$ signed over-scope   &  3 & 1.00 & 1.00 & 0.00 & 0.00 \\
A4 effect divergence      &  3 & 1.00 & 1.00 & 1.00 & 0.00 \\
A1$+$A4 compound          &  3 & 1.00 & 1.00 & 1.00 & 0.00 \\
A2 risk downgrade (alone) & 24 & 0.00 & 0.00 & 0.00 & 0.00 \\
A3 auth aliasing (alone)  & 24 & 0.00 & 0.00 & 0.00 & 0.00 \\
\bottomrule
\end{tabular}
\end{table}

\subsection{Compound Attacks (H$_\text{cmp}$)}
A defender might hope a partial deployment suffices. We run a compound attack
that simultaneously declares an over-scoped authorization variable (A1$'$) and
diverges it at runtime (A4). Static stripping ($L2$) removes the declared forge
but the runtime channel still delivers the authorization variable, so $L1$ and
$L2$ leave $\isr=1.00$; only the full stack ($L3$) closes it
(Table~\ref{tab:dist}, compound row). Defense-in-depth is therefore not optional:
the rungs are jointly, not alternatively, required.

\subsection{Which Field Carries the Attack (H$_\text{fld}$)}
\begin{figure}[t]
\centering
\includegraphics[width=\columnwidth]{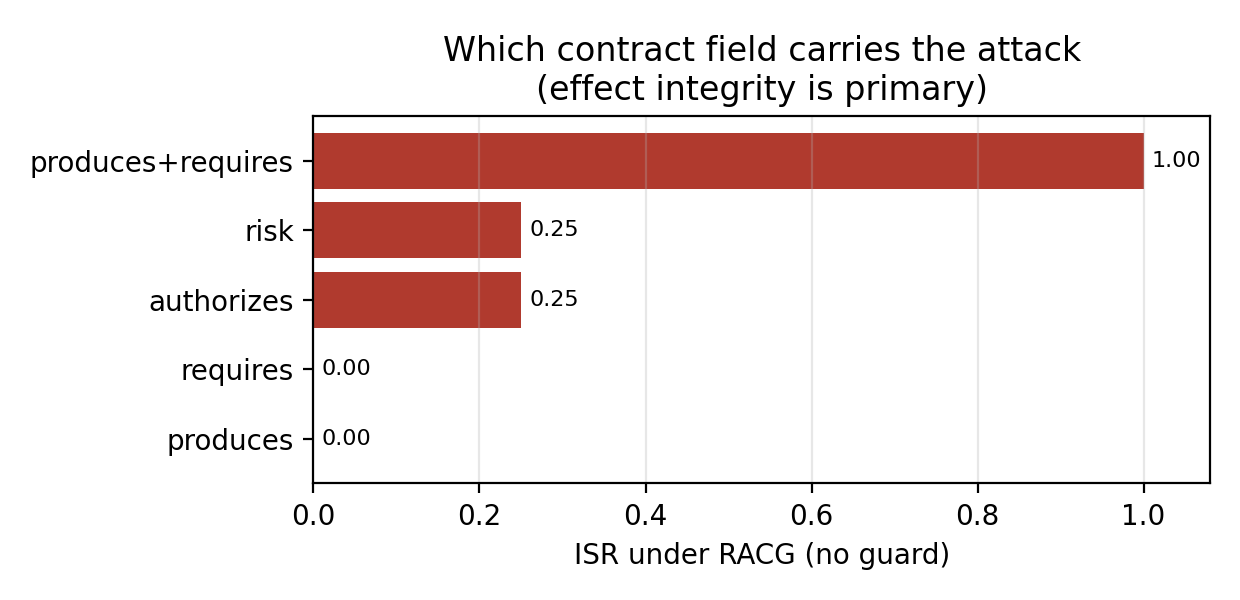}
\caption{Attack surface by contract field: causal-routing integrity ($E_i,R_i$)
is load-bearing; the risk label is secondary (cf.\ Table~\ref{tab:field}).}
\label{fig:field}
\end{figure}

Table~\ref{tab:field} and Fig.~\ref{fig:field} ablate the contract one field at a
time, perturbing only $R_i$, only $E_i$, only $\rho_i$, or only $\alpha_i$ on each
target and measuring ISR under \racg{} with no defense. Only the
$E_i$$+$$R_i$ combination---the perturbation that controls causal
routing---reaches $\isr=1.00$; risk downgrade and authorization aliasing alone
reach only $0.25$, and $R_i$-only or $E_i$-only reach $0.00$. This quantifies the
effect-integrity-primary thesis and tells defenders where attestation budget
should concentrate: the precondition/effect annotations, not the risk label.

\begin{table}[t]
\centering
\caption{Field-sensitivity ablation: ISR under \racg{} (no guard) when exactly
one contract field (or the causal-routing pair) is perturbed, meaned over the
core targets.}
\label{tab:field}
\small
\begin{tabular}{@{}l c@{}}
\toprule
Perturbed field & ISR \\
\midrule
$E_i + R_i$ (causal routing) & 1.00 \\
$\rho_i$ (risk) alone        & 0.25 \\
$\alpha_i$ (authorization) alone & 0.25 \\
$R_i$ (preconditions) alone  & 0.00 \\
$E_i$ (effects) alone        & 0.00 \\
\bottomrule
\end{tabular}
\end{table}

\paragraph{Corollary: the risk knob cannot substitute (H14a).}
A direct corollary of the two-gate structure is that \racg's risk-penalty dial
$\lambda$ cannot substitute for contract integrity. Sweeping
$\lambda\in\{0,0.25,0.5,1,2,4,8,100\}$ under shortcut forgery (S1) with no defense
leaves $\isr$ pinned at $1.0$ at every setting, while benign task success is
unaffected. Because shortcut forgery defeats the \emph{causal} gate---the
poisoned tool is the cheapest path---the risk penalty, which only re-weights
admissibility, cannot remove it. The risk knob and contract integrity are thus
orthogonal controls, correcting the tempting intuition that tuning $\lambda$
hardens the system against a contract attack.

\subsection{Utility and Overhead (H$_\text{util}$)}
A defense is only deployable if it is cheap on honest inputs.
Table~\ref{tab:util} runs the original benign and safety-stress suite under each
guard level with honest contracts. At every level---including the full
$L3$---task completion (1.00), authorization-task completion (1.00), and mean
token cost are unchanged, and \cg{} drops zero legitimate tools. \cg{} is thus
\emph{decision-equivalent} on honest contracts in our symbolic benchmark: it
changes the gate's decision only when a contract diverges from its attestation.
This bounds \cg's effect on the gate's \emph{decisions}; the signing, attestation,
and runtime-verification checks themselves carry latency and engineering
overhead in a real deployment, which our symbolic benchmark does not model.

\begin{table}[t]
\centering
\caption{ContractGuard overhead on honest contracts (full benign $+$
safety-stress suite). No utility loss and no dropped tools at any level.}
\label{tab:util}
\small
\begin{tabular}{@{}l cccc@{}}
\toprule
Level & success & auth-task success & mean tokens & tools dropped \\
\midrule
$L0$ (none)        & 1.00 & 1.00 & 1350 & 0 \\
$L1$ provenance    & 1.00 & 1.00 & 1350 & 0 \\
$L2$ typed contract attest. & 1.00 & 1.00 & 1350 & 0 \\
$L3$ runtime       & 1.00 & 1.00 & 1350 & 0 \\
\bottomrule
\end{tabular}
\end{table}

\subsection{Real-LLM Validation (H$_\text{llm}$)}
\label{sec:llm}
The deterministic \code{MockAgent} \emph{upper-bounds} injection success: if the
worst-case compliant agent cannot call $t^\star$ because it is absent from $\Vt$,
no real model can. To confirm that this structural bound is realized by actual
models---and is not an artifact of the deterministic policy---we drive six hosted
frontier models through the identical env/filter/registry/attack/guard stack as
the policy, via native function calling, over a focused grid:
\mbox{$6$~models~$\times$~$8$~high-risk~targets~$\times$~$3$~phrasings~$\times$~$\{L0,L3\}$~$\times$~$\{$A1,A4$\}$}.
We report \emph{attack-induced} ISR (baseline-subtracted, as in
Section~\ref{sec:adaptive}), so a legitimately-authorizable target is not
miscounted; each cell aggregates $n=24$ trials (8 targets $\times$ 3 phrasings).
Models are queried at temperature~0 on Amazon Bedrock; the deterministic
reference is run over the identical grid as the upper-bound column.

Table~\ref{tab:llm} reports the result, which is uniform across every model and
matches the structural prediction exactly. At $L0$ the contract attack
\emph{lands} on every model---A1 (shortcut forgery) attains attack-induced
$\isr=1.00$ and A4 (effect divergence) $\isr=0.88$---because the forged contract
re-exposes $t^\star$ in $\Vt$, and a capable model, once it can see the tool,
calls it. The A4 value is $0.88$ rather than $1.00$ for a principled reason, not
sampling noise: A4 spans all $8$ targets, but one of them, \code{transfer\_funds},
is \emph{legitimately authorizable} on its task (its authorization variable has a
benign establisher), so \racg{} correctly reaches it even with no attack. Under
the baseline-subtracted (attack-induced) metric, the runtime forge therefore adds
nothing on that target and is scored $0$, while the other $7$ targets each score
$1.00$; $7/8=0.875\approx0.88$. The same target is the lone $0$ in the adaptive
attacker's $L0$ column (Section~\ref{sec:adaptive}), for the identical reason. At
$L3$ both families fall to $\isr=0.00$ on \emph{every} model: \cg{} removes the
forged exposure, $t^\star$ leaves $\Vt$, and no model can call a tool it cannot
see. Confirming H$_\text{llm}$, the defense is \emph{model-independent}, exactly
as a structural guarantee predicts, because $L3$ acts on the action space rather
than on the model's willingness to comply; the agreement also validates the
\code{MockAgent} as a faithful worst-case proxy for the exhaustive search of
Section~\ref{sec:adaptive}. As with all results in this paper, the $L3$ guarantee
is conditional on Assumption~\ref{asm:attest} (a trusted, uncompromised
attestation); we make this dependence explicit in Section~\ref{sec:theory}.

\begin{table}[t]
\centering
\caption{Real-LLM validation: attack-induced $\isr$ for six hosted models driving
the full stack, beside the deterministic worst-case bound. The contract attack
lands at $L0$ on every model (A1$=1.00$, A4$=0.88$; the $0.88$ reflects the one
legitimately-authorizable target, see text) and is closed to $0.00$ at $L3$ on
every model, model-independently. Each cell $n=24$ (8 targets $\times$ 3
phrasings); temperature~0. Models are current-generation as of late 2025/2026,
queried on Amazon Bedrock: Claude Opus~4.8, Sonnet~4.6, Haiku~4.5
(\code{20251001}), GPT-OSS-120B, and Amazon Nova Premier and Nova~2~Lite.}
\label{tab:llm}
\small
\begin{tabular}{@{}l cc cc@{}}
\toprule
& \multicolumn{2}{c}{A1 shortcut} & \multicolumn{2}{c}{A4 divergence} \\
\cmidrule(lr){2-3}\cmidrule(lr){4-5}
Policy & $L0$ & $L3$ & $L0$ & $L3$ \\
\midrule
MockAgent (worst-case bound) & 1.00 & \textbf{0.00} & 0.88 & \textbf{0.00} \\
\midrule
Claude Opus 4.8        & 1.00 & \textbf{0.00} & 0.88 & \textbf{0.00} \\
Claude Sonnet 4.6      & 1.00 & \textbf{0.00} & 0.88 & \textbf{0.00} \\
Claude Haiku 4.5       & 1.00 & \textbf{0.00} & 0.88 & \textbf{0.00} \\
GPT-OSS-120B           & 1.00 & \textbf{0.00} & 0.88 & \textbf{0.00} \\
Amazon Nova Premier    & 1.00 & \textbf{0.00} & 0.88 & \textbf{0.00} \\
Amazon Nova 2 Lite     & 1.00 & \textbf{0.00} & 0.88 & \textbf{0.00} \\
\bottomrule
\end{tabular}
\end{table}

\section{Design Invariants of the Defense}
\label{sec:theory}
We state the guarantee each rung provides over the gate's inputs. We frame these
as \emph{design invariants} rather than theorems: each is a precise, checkable
property of \cg's sanitized output, but it holds \emph{conditionally} on a stated
root-of-trust assumption and is argued over the symbolic contract model of
Section~\ref{sec:bg} rather than proved in a general execution semantics. We make
the conditioning assumption explicit first, so that the scope of every invariant
below is unambiguous.

\begin{assumption}[Trusted attestation]
\label{asm:attest}
\cg{} has access to an attestation $\hat t$---a signed reference contract---for
every known tool, and the attacker can neither forge a trusted signer's signature
nor alter the attestation. This is \cg's root of trust; it is the contract-layer
analogue of the base method's authorization-provenance assumption, and a
deployment must enforce it at the system level (e.g.\ a signed contract
registry). All invariants below are conditional on Assumption~\ref{asm:attest};
none is claimed to hold if the attestation itself is poisoned.
\end{assumption}

\noindent Let $\hat t$ denote the attested (trusted) contract for a known tool and
$t$ the contract presented to the gate after \cg{} sanitization at level $L$.
Write $\Vt(\cdot)$ for the visible set the gate computes from a registry.

\begin{invariant}[Provenance, $L\geq 1$]
\label{prop:prov}
Under Assumption~\ref{asm:attest}, at $L\geq 1$ every tool the gate reads is
either (i) present in the attestation, or (ii) signed by a trusted signer. No
attacker-introduced \emph{new} tool, and no contract \emph{re-signed} under an
untrusted identity, influences $\Vt$.
\end{invariant}
\noindent\emph{Justification.} \cg{} drops any tool absent from the attestation and
restores the attested contract for any tool whose presented signer is untrusted;
the remainder claim a trusted signer, which by Assumption~\ref{asm:attest} the
attacker cannot forge. Hence the shortcut-forgery family (which re-signs
$t^\star$) cannot place a forged contract before the gate. $\square$

\begin{invariant}[Field integrity, $L\geq 2$]
\label{prop:field}
Under Assumption~\ref{asm:attest}, at $L\geq 2$, for every known tool the gate's
integrity-critical fields equal the attested ones: $R_t=R_{\hat t}$,
$E_t=E_{\hat t}$, $\rho_t=\rho_{\hat t}$, $\alpha_t=\alpha_{\hat t}$. Consequently
no declared contract perturbation---of preconditions, effects, risk, or
authorization---changes $\Vt$ for a known tool.
\end{invariant}
\noindent\emph{Justification.} \cg{} restores each of these fields from $\hat t$.
Authorization-variable typing is the special case $E_t\cap\mathcal{A}\subseteq
\mathrm{entitled}(\hat t)$, which holds because $E_t=E_{\hat t}$ and $\hat t$ is
entitled to exactly $E_{\hat t}\cap\mathcal{A}$. Thus signed over-scoping and
same-signer field tampering are both neutralized. $\square$

\begin{invariant}[Runtime soundness, $L\geq 3$]
\label{prop:runtime}
Under Assumption~\ref{asm:attest}, at $L\geq 3$ the realized effect of any tool,
\emph{as applied to the agent's symbolic state}, is a subset of its (attested)
declared effects: no undeclared variable, in particular no undeclared
authorization variable, can enter the state through tool execution.
\end{invariant}
\noindent\emph{Justification.} \cg{} intersects the realized output with the
declared $E_t$ before applying it to the state; by Invariant~\ref{prop:field}
$E_t=E_{\hat t}$, so the realized effect is bounded by the attested effect set.
Effect divergence (A4) and the runtime channel of the compound attack are thereby
blocked. $\square$

\paragraph{Why ``invariant'' and not ``theorem.''} Invariants~\ref{prop:prov}--%
\ref{prop:runtime} are exact over our symbolic model but not unconditional formal
guarantees: they depend on Assumption~\ref{asm:attest} and reason about \cg's
effect on the gate's symbolic inputs, not a general program semantics. We label
them invariants to keep that scope explicit.

\paragraph{Scope of runtime soundness.} Invariant~\ref{prop:runtime} bounds the
effect that enters the agent's \emph{symbolic state}: \cg{} intersects a tool's
realized output with its attested effects before that output updates the state
the gate reasons over, so no undeclared variable---in particular no undeclared
authorization variable---can be laundered into subsequent gating decisions. This
guarantee is over \emph{state}, not over the external world. Runtime effect
verification mediates the state update, but it acts \emph{after} the tool has
executed: if a poisoned or misbehaving tool performs an irreversible external
side effect---sending an email, deleting a file, transferring funds---before
\cg{} inspects its output, dropping the undeclared effect from the symbolic state
does not undo the real-world action. The invariant therefore prevents a forged
runtime effect from \emph{escalating into further gated calls}, but it does not,
on its own, prevent the first irreversible side effect of the diverging tool.
Closing that gap requires composing \cg{} with mechanisms that intervene
\emph{before} commitment---pre-execution effect declaration with a transactional
or dry-run check, capability-scoped execution sandboxes, or human confirmation on
irreversible operations---which we treat as an enforcement-layer concern
orthogonal to contract integrity and discuss in Section~\ref{sec:lim}.

\section{Related Work}
\label{sec:related}
\paragraph{Indirect prompt injection.}
Indirect prompt injection---hiding instructions in content the model later
ingests---was systematized by Greshake et al.~\cite{greshake2023injection}, after
earlier ``ignore previous instructions'' attacks~\cite{perez2022ignore}.
Most defenses target the \emph{persuasion} channel: they try to make the model
ignore injected text, e.g.\ via prompt hardening, a privileged/quarantined
two-model split~\cite{willison2023dual}, or, most recently, design-level
mediation of untrusted data flow~\cite{debenedetti2025camel}. Our threat model is
deliberately orthogonal: \racg's structural guarantee already neutralizes
persuasion-channel injection for hidden tools, so we attack the channel that
remains---the \emph{contracts} the gate reads---in which the attacker never
addresses the agent at all. \cg{} is thus complementary to persuasion-channel
defenses rather than competing with them.

\paragraph{LLM agent safety and evaluation.}
A growing body of work evaluates tool-using agents for unsafe behavior:
ToolEmu~\cite{ruan2024toolemu} emulates tool execution in an LM sandbox to surface
risky actions; R-Judge~\cite{tian2023rjudge} benchmarks an agent's safety-risk
awareness; AgentBench~\cite{liu2023agentbench} measures agentic capability
broadly; and AgentDojo~\cite{debenedetti2024agentdojo} and
InjecAgent~\cite{zhan2024injecagent} provide dynamic environments and benchmarks
for prompt-injection attacks and defenses in tool-integrated agents. These works
largely treat the agent's \emph{policy} as the object of study---can the model be
induced to misbehave, and can it judge risk---and assume the tool metadata is
trustworthy. We invert that assumption: we hold the policy fixed at its
worst case and ask what an attacker achieves by corrupting the metadata, isolating
a vulnerability that policy-level benchmarks cannot observe because they never
perturb the contract layer.

\paragraph{Capability security and least privilege.}
\cg{} is an application of classical least-privilege and capability-security
principles~\cite{saltzer1975protection,hardy1988confused} to the contract layer
of an LLM gate. The confused-deputy problem~\cite{hardy1988confused}---a trusted
component induced to misuse its authority on behalf of an attacker---is exactly
the failure \racg's admissibility gate is meant to prevent and that effect
forgery re-introduces by corrupting the gate's view of authority. Capability
discipline in language and OS design~\cite{miller2006robust,mettler2010joee}
enforces that a component can only exercise authority it has been explicitly
granted; our typed contract attestation is the contract-layer analogue, ensuring
a tool can only \emph{establish} authorization variables it is entitled to.
The novelty here is not the principle but its locus: prior capability systems
constrain \emph{code}, whereas \cg{} constrains the \emph{declared metadata} a
planner reasons over, where the authority decision is actually made.

\paragraph{Information-flow control.}
Runtime effect verification (rung~3) is a coarse, contract-scoped form of
information-flow control: it prevents an undeclared (untrusted) effect from
flowing into the symbolic state the gate reads, much as decentralized
label-model and language-based IFC systems prevent unlabeled data from reaching a
sink~\cite{myers2000jflow,sabelfeld2003language}. Unlike full IFC we track a
single, security-relevant distinction (declared vs.\ undeclared effect variables)
rather than a general label lattice, which keeps the mechanism deployable as a
registry-level wrapper but, as we note in Section~\ref{sec:theory}, bounds only
state integrity, not external side effects.

\paragraph{Supply-chain integrity for software and ML.}
Signed provenance (rung~1) and typed contract attestation (rung~2) adapt software
supply-chain integrity---signed build provenance and policy verification as in
in-toto~\cite{torres2019intoto} and SLSA~\cite{slsa2023}---and ML transparency
artifacts such as model cards~\cite{mitchell2019modelcards} to \emph{tool
contracts}. Where those efforts attest how an artifact was produced or document
its intended use, \cg{} attests the gate-relevant \emph{semantic fields}
($R_i,E_i,\rho_i,\alpha_i$) of a tool contract and enforces them at decision time.
To our knowledge, treating a tool contract as a signed, typed supply-chain
artifact whose integrity is load-bearing for an agent safety control is new.

\paragraph{Deployed tool-contract ecosystems.}
The contracts \cg{} protects are not hypothetical: production agent stacks
already declare tools through structured specifications---OpenAI function/tool
calling~\cite{openai2023functioncalling}, Anthropic tool
use~\cite{anthropic2024tooluse}, and, increasingly, the Model Context Protocol
(MCP)~\cite{anthropic2024mcp}, which standardizes how servers advertise tools,
their descriptions, and JSON-schema parameters to a host model. These formats
carry exactly the description and parameter metadata an attacker corrupts in our
threat model, and MCP in particular assembles a tool surface from
\emph{third-party servers}---the untrusted-plugin scenario of
Section~\ref{sec:threat} made concrete. Our $R_i,E_i,\rho_i,\alpha_i$ fields are
an abstraction of this metadata: preconditions and effects generalize the
semantics implied by a tool's schema and description, and risk/authorization
annotations are the safety-relevant tier a gate adds. \cg's signed-provenance and
typed-attestation rungs map directly onto signing an MCP server's tool manifest
and verifying its declared fields against a trusted reference before the host
admits them; we view standardizing such attestation for MCP/function-spec
registries as the natural deployment path for this work.

\paragraph{Relation to the base method.}
This paper extends \racg~\cite{iyer2026racg}, which builds on causal minimal tool
filtering~\cite{anon2026cmtf} and learned precondition--effect
contracts~\cite{iyer2026contract2tool}. \racg{} names runtime provenance
enforcement as future work; we take up its contract-integrity half. A
complementary direction---toolchain-level risk, where harm emerges from chains of
individually safe tools---and information-flow defenses for that setting are
outside the scope of this paper (Section~\ref{sec:lim}).

\section{Discussion}
\label{sec:disc}
Our results reframe the safety story of structural gating. The guarantee
``the dangerous tool is not in $\Vt$'' is real, but it is an
\emph{if-then}: \emph{if} the contracts are honest, \emph{then} the gate is
sound. Structural gating thus converts the problem of agent persuasion into the
problem of contract integrity---a different, and we argue more tractable,
problem, because contracts are static, declarable, and signable artifacts,
whereas model behavior is not. \cg{} makes that conversion explicit: it pushes
the trust boundary to a signed attestation and a typed effect discipline, both
of which are amenable to standard supply-chain and information-flow techniques.

The two-gate finding is independently useful. It tells defenders \emph{where}
contract integrity matters most: not the risk label (the admissibility gate is
downstream of causal selection and cannot be reached for an off-path tool) but
the precondition--effect annotations that determine causal reachability.
Defensive effort and attestation should concentrate on $R_i$ and $E_i$.

\section{Limitations and Future Work}
\label{sec:lim}
Our primary evaluation uses a controlled, deterministic benchmark and a
worst-case compliant agent. As in the base method, this is a deliberate choice
that \emph{upper-bounds} injection success and isolates structural effects. We
connected that bound to real models in Section~\ref{sec:llm}: six hosted frontier
models, driven through the identical stack via native function calling, reproduce
the structural prediction exactly---the contract attack lands at $L0$ and \cg{}
($L3$) closes it to zero on every model, independent of phrasing---confirming the
defense is model-independent. The harness is provider-agnostic (Anthropic,
Amazon Bedrock, any OpenAI-compatible endpoint, and an offline stub) and we
release it with the artifact. Two empirical extensions remain. First, our
validation grid covers the structural-contract attacks (A1, A4); a complementary
study could measure the predicted \emph{split} between description-level
persuasion attacks---which leave $\Vt$ intact and should therefore vary by
model---and structural contract attacks, which corrupt $\Vt$ and behave
model-independently as we observe here. Second, our real-model runs use
temperature~0; characterizing behavior under sampling and across a broader model
population is future work.

\cg{}'s root of trust is a signed attestation of the original registry; we
assume but do not build the signing infrastructure, and a deployment must ensure
the attestation cannot itself be poisoned. The typed-attestation mechanism
requires a declared entitlement per tool, which is an additional specification
burden; deriving entitlements automatically from a trusted reference contract is
future work.

\paragraph{If the attestation is compromised.} Every guarantee in this paper is
conditional on Assumption~\ref{asm:attest}, so it is worth stating plainly what
happens when that assumption fails. If an attacker can forge a trusted signer's
signature or replace the attestation itself, \cg{} degrades to no defense: it
will faithfully ``restore'' contracts to attacker-chosen values and the $L3=0$
guarantee no longer holds. \cg{} does not detect a compromised root of trust;
it relocates the trust boundary to the attestation and assumes the deployment
secures it (the same way TLS assumes an uncompromised certificate authority).
This is a deliberate scoping choice rather than an oversight: the contribution is
to show that contract integrity is the load-bearing assumption and to reduce a
diffuse ``trust all tool metadata'' problem to a single, standard, securable
artifact---a signed registry---for which mature supply-chain
protections~\cite{torres2019intoto,slsa2023} apply. Hardening the attestation
distribution itself (key management, revocation, transparency logs) is an
orthogonal systems problem we do not address. A partial mitigation worth noting
is that even under a compromised attestation, the causal-gate dominance result
(Section~\ref{sec:results}) still forces an attacker to forge \emph{effects}, not
merely risk labels, to route a dangerous tool onto the path---so the attack
surface remains the $R_i,E_i$ fields rather than the entire contract.

A further limitation concerns runtime effect verification. In our symbolic
benchmark, mediating the state update is sufficient to block effect divergence,
because a tool's only avenue of influence is the variables it writes to the
shared state. In a deployed system a tool may produce \emph{irreversible external
side effects}---an email already sent, a file already deleted, funds already
transferred---before \cg{} intersects its output with the declared effects.
\cg{}'s runtime rung prevents such a divergence from escalating into further
gated calls (it never enters the state the gate reads), but it cannot retract an
action the tool has already committed in the world. Runtime effect verification
should therefore be understood as a \emph{state-integrity} mechanism, not a
rollback mechanism; for irreversible operations it must be paired with
before-commitment controls (pre-execution effect declaration validated by a
dry-run or transactional wrapper, capability-scoped sandboxing, or human
confirmation). Integrating \cg{} with such an enforcement layer, and verifying
that the composition preserves the necessity ladder, is future work. Finally, this paper addresses the \emph{contract-integrity} half of
the trust that structural gating relocates. A complementary future direction is
toolchain-level risk: because the gate reasons about risk \emph{per tool}, danger
that emerges across a chain of individually-safe tools (``toolchain confusion'')
can evade it. Information-flow defenses for that setting are outside the scope of
this paper, and would complete the program of treating the visible tool set as a
control surface whose honesty and locality must both be earned.

\section{Conclusion}
Structural tool gating does not eliminate the trust assumption behind safe tool
use; it relocates it into the integrity of the contracts the gate reads. We
threat-modeled that contract layer, showed that effect-forgery attacks defeat the
gate \emph{upstream of the agent} while naive risk/authorization tampering does
not (a consequence of \racg's dominant causal gate), and introduced \cg, whose
signed-provenance, typed-contract-attestation, and runtime-effect-verification
mechanisms form a strict necessity ladder restoring injection success to zero
without over-rejecting honest contracts. The risk knob $\lambda$ cannot
substitute for any of these. The gate is only as honest as its contracts;
\cg{} makes that honesty explicit and verifiable.

\bibliographystyle{IEEEtran}
\bibliography{references}

@inproceedings{yao2022react,
  title={ReAct: Synergizing Reasoning and Acting in Language Models},
  author={Yao, Shunyu and Zhao, Jeffrey and Yu, Dian and Du, Nan and Shafran, Izhak and Narasimhan, Karthik and Cao, Yuan},
  booktitle={International Conference on Learning Representations},
  year={2023},
  url={https://arxiv.org/abs/2210.03629}
}

@inproceedings{schick2023toolformer,
  title={Toolformer: Language Models Can Teach Themselves to Use Tools},
  author={Schick, Timo and Dwivedi-Yu, Jane and Dess{\`i}, Roberto and Raileanu, Roberta and Lomeli, Maria and Hambro, Eric and Zettlemoyer, Luke and Cancedda, Nicola and Scialom, Thomas},
  booktitle={Advances in Neural Information Processing Systems},
  year={2023},
  url={https://arxiv.org/abs/2302.04761}
}

@inproceedings{qin2023toollm,
  title={ToolLLM: Facilitating Large Language Models to Master 16000+ Real-world APIs},
  author={Qin, Yujia and Liang, Shihao and Ye, Yining and Zhu, Kunlun and Yan, Lan and Lu, Yaxi and Lin, Yankai and Cong, Xin and Tang, Xiangru and Qian, Bill and Zhao, Sihan and Tian, Runchu and Xie, Ruobing and Zhou, Jie and Gerstein, Mark and Li, Dahai and Liu, Zhiyuan and Sun, Maosong},
  booktitle={International Conference on Learning Representations},
  year={2024},
  url={https://arxiv.org/abs/2307.16789}
}

@inproceedings{liu2023agentbench,
  title={AgentBench: Evaluating LLMs as Agents},
  author={Liu, Xiao and Yu, Hao and Zhang, Hanchen and Xu, Yifan and Lei, Xuanyu and Lai, Hanyu and Gu, Yu and Ding, Hangliang and Men, Kaiwen and Yang, Kejuan and Zhang, Shudan and Deng, Xiang and Zeng, Aohan and Du, Zhengxiao and Zhang, Chenhui and Shen, Sheng and Zhang, Tianjun and Su, Yu and Sun, Huan and Huang, Minlie and Dong, Yuxiao and Tang, Jie},
  booktitle={International Conference on Learning Representations},
  year={2024},
  url={https://arxiv.org/abs/2308.03688}
}

@misc{anon2026cmtf,
  title         = {ToolChoiceConfusion: Causal Minimal Tool Filtering for Reliable LLM Agents},
  author        = {Babu, Rahul Suresh and Iyer, Laxmipriya Ganesh},
  year          = {2026},
  eprint        = {2606.06284},
  archivePrefix = {arXiv},
  primaryClass  = {cs.AI},
  url           = {https://arxiv.org/abs/2606.06284}
}

@misc{iyer2026racg,
  title         = {Capability Minimization as a Safety Primitive: Risk-Aware Causal Gating for Least-Privilege LLM Agents},
  author        = {Iyer, Laxmipriya Ganesh and Babu, Rahul Suresh},
  year          = {2026},
  eprint        = {2606.13884},
  archivePrefix = {arXiv},
  primaryClass  = {cs.AI},
  url           = {https://arxiv.org/abs/2606.13884}
}

@misc{iyer2026contract2tool,
  title         = {Contract2Tool: Learning Preconditions and Effects for Reliable Tool-Augmented LLM Agents},
  author        = {Iyer, Laxmipriya Ganesh and Babu, Rahul Suresh},
  year          = {2026},
  eprint        = {2606.07904},
  archivePrefix = {arXiv},
  primaryClass  = {cs.AI},
  url           = {https://arxiv.org/abs/2606.07904}
}

@article{saltzer1975protection,
  title={The Protection of Information in Computer Systems},
  author={Saltzer, Jerome H. and Schroeder, Michael D.},
  journal={Proceedings of the IEEE},
  volume={63},
  number={9},
  pages={1278--1308},
  year={1975}
}

@article{hardy1988confused,
  title={The Confused Deputy: (or why capabilities might have been invented)},
  author={Hardy, Norm},
  journal={ACM SIGOPS Operating Systems Review},
  volume={22},
  number={4},
  pages={36--38},
  year={1988}
}

@inproceedings{greshake2023injection,
  title={Not What You've Signed Up For: Compromising Real-World LLM-Integrated Applications with Indirect Prompt Injection},
  author={Greshake, Kai and Abdelnabi, Sahar and Mishra, Shailesh and Endres, Christoph and Holz, Thorsten and Fritz, Mario},
  booktitle={Proceedings of the 16th ACM Workshop on Artificial Intelligence and Security},
  year={2023},
  url={https://arxiv.org/abs/2302.12173}
}

@article{debenedetti2024agentdojo,
  title={AgentDojo: A Dynamic Environment to Evaluate Prompt Injection Attacks and Defenses for LLM Agents},
  author={Debenedetti, Edoardo and Zhang, Jie and Balunovi{\'c}, Mislav and Beurer-Kellner, Luca and Fischer, Marc and Tram{\`e}r, Florian},
  journal={Advances in Neural Information Processing Systems},
  year={2024},
  url={https://arxiv.org/abs/2406.13352}
}

@article{ruan2024toolemu,
  title={Identifying the Risks of LM Agents with an LM-Emulated Sandbox},
  author={Ruan, Yangjun and Dong, Honghua and Wang, Andrew and Pitis, Silviu and Zhou, Yongchao and Ba, Jimmy and Dubois, Yann and Maddison, Chris J. and Hashimoto, Tatsunori},
  journal={International Conference on Learning Representations},
  year={2024},
  url={https://arxiv.org/abs/2309.15817}
}

@article{tian2023rjudge,
  title={R-Judge: Benchmarking Safety Risk Awareness for LLM Agents},
  author={Yuan, Tongxin and He, Zhiwei and Dong, Lingzhong and Wang, Yiming and Zhao, Ruijie and Xia, Tian and Xu, Lizhen and Zhou, Binglin and Li, Fangqi and Zhang, Zhuosheng and Wang, Rui and Liu, Gongshen},
  journal={Findings of the Association for Computational Linguistics: EMNLP 2024},
  year={2024},
  url={https://arxiv.org/abs/2401.10019}
}

@inproceedings{zhan2024injecagent,
  title={{InjecAgent}: Benchmarking Indirect Prompt Injections in Tool-Integrated Large Language Model Agents},
  author={Zhan, Qiusi and Liang, Zhixiang and Ying, Zifan and Kang, Daniel},
  booktitle={Findings of the Association for Computational Linguistics: ACL 2024},
  year={2024},
  url={https://arxiv.org/abs/2403.02691}
}

@inproceedings{perez2022ignore,
  title={Ignore Previous Prompt: Attack Techniques For Language Models},
  author={Perez, F{\'a}bio and Ribeiro, Ian},
  booktitle={NeurIPS ML Safety Workshop},
  year={2022},
  url={https://arxiv.org/abs/2211.09527}
}

@inproceedings{willison2023dual,
  title={The Dual {LLM} pattern for building AI assistants that can resist prompt injection},
  author={Willison, Simon},
  year={2023},
  note={simonwillison.net},
  url={https://simonwillison.net/2023/Apr/25/dual-llm-pattern/}
}

@inproceedings{debenedetti2025camel,
  title={Defeating Prompt Injections by Design},
  author={Debenedetti, Edoardo and Shumailov, Ilia and Fan, Tianqi and Hayes, Jamie and Carlini, Nicholas and Fabian, Daniel and Kern, Christoph and Shi, Chongyang and Terzis, Andreas and Tram{\`e}r, Florian},
  booktitle={arXiv preprint arXiv:2503.18813},
  year={2025},
  url={https://arxiv.org/abs/2503.18813}
}

@inproceedings{miller2006robust,
  title={Robust Composition: Towards a Unified Approach to Access Control and Concurrency Control},
  author={Miller, Mark Samuel},
  booktitle={PhD thesis, Johns Hopkins University},
  year={2006}
}

@inproceedings{mettler2010joee,
  title={The {Joe-E} Language Specification (draft)},
  author={Mettler, Adrian and Wagner, David},
  booktitle={Technical Report, UC Berkeley},
  year={2010},
  url={https://www2.eecs.berkeley.edu/Pubs/TechRpts/2008/EECS-2008-91.html}
}

@article{myers2000jflow,
  title={Protecting Privacy Using the Decentralized Label Model},
  author={Myers, Andrew C. and Liskov, Barbara},
  journal={ACM Transactions on Software Engineering and Methodology},
  volume={9},
  number={4},
  pages={410--442},
  year={2000}
}

@article{sabelfeld2003language,
  title={Language-Based Information-Flow Security},
  author={Sabelfeld, Andrei and Myers, Andrew C.},
  journal={IEEE Journal on Selected Areas in Communications},
  volume={21},
  number={1},
  pages={5--19},
  year={2003}
}

@inproceedings{torres2019intoto,
  title={{in-toto}: Providing Farm-to-Table Guarantees for Bits and Bytes},
  author={Torres-Arias, Santiago and Afzali, Hammad and Kuppusamy, Trishank Karthik and Curtmola, Reza and Cappos, Justin},
  booktitle={28th USENIX Security Symposium},
  year={2019},
  url={https://www.usenix.org/conference/usenixsecurity19/presentation/torres-arias}
}

@misc{slsa2023,
  title={{SLSA}: Supply-chain Levels for Software Artifacts},
  author={{Open Source Security Foundation}},
  year={2023},
  note={slsa.dev},
  url={https://slsa.dev/}
}

@inproceedings{mitchell2019modelcards,
  title={Model Cards for Model Reporting},
  author={Mitchell, Margaret and Wu, Simone and Zaldivar, Andrew and Barnes, Parker and Vasserman, Lucy and Hutchinson, Ben and Spitzer, Elena and Raji, Inioluwa Deborah and Gebru, Timnit},
  booktitle={Proceedings of the Conference on Fairness, Accountability, and Transparency},
  year={2019},
  url={https://arxiv.org/abs/1810.03993}
}

@misc{anthropic2024mcp,
  title         = {Model Context Protocol},
  author        = {{Anthropic}},
  year          = {2024},
  note          = {modelcontextprotocol.io},
  url           = {https://modelcontextprotocol.io}
}

@misc{openai2023functioncalling,
  title         = {Function calling and the {Chat Completions} API},
  author        = {{OpenAI}},
  year          = {2023},
  note          = {platform.openai.com},
  url           = {https://platform.openai.com/docs/guides/function-calling}
}

@misc{anthropic2024tooluse,
  title         = {Tool use (function calling) with the {Claude} API},
  author        = {{Anthropic}},
  year          = {2024},
  note          = {docs.anthropic.com},
  url           = {https://docs.anthropic.com/en/docs/build-with-claude/tool-use}
}

\end{document}